\documentclass[a4paper, 11pt]{article}
\makeatletter
	\def\ps@pprintTitle{%
 	\let\@oddhead\@empty
	\let\@evenhead\@empty
	\def\@oddfoot{\centerline{\thepage}}%
	\let\@evenfoot\@oddfoot}
\makeatother

\usepackage{etoolbox}
\patchcmd{\MaketitleBox}{\footnotesize\itshape\elsaddress\par\vskip36pt}{\footnotesize\itshape\elsaddress\par\parbox[b][36pt]{\linewidth}{\vfill\hfill\textnormal{\today}\hfill\null\vfill}}{}{}%
\patchcmd{\pprintMaketitle}{\footnotesize\itshape\elsaddress\par\vskip36pt}{\footnotesize\itshape\elsaddress\par\parbox[b][36pt]{\linewidth}{\vfill\hfill\textnormal{\today}\hfill\null\vfill}}{}{}%

\usepackage{geometry}
\usepackage{latexsym,bm}
\usepackage[tbtags]{amsmath}
\usepackage{amssymb}
\usepackage{graphicx}
\usepackage{setspace}
\usepackage{booktabs}
\usepackage{multirow}
\usepackage{caption}
\usepackage{subcaption}
\usepackage{placeins}
\usepackage{float}
\usepackage{romannum}
\usepackage{hyperref}
\usepackage[utf8]{inputenc}
\usepackage{CJKutf8}
\usepackage[T1]{fontenc}
\usepackage{array}
\usepackage{varwidth}
\usepackage{rotating}
\usepackage{tabularx}
\usepackage{makecell}     
\usepackage{adjustbox}    
\newtheorem{problem}{Problem}
\usepackage{authblk}
\usepackage{abstract}
\usepackage{titlesec}
\usepackage{siunitx}

\titleformat{\section}[block]{\normalfont\bfseries\centering}{\Roman{section}.}{0.5em}{}
\titleformat{\subsection}[block]{\normalfont\bfseries}{\thesection.\arabic{subsection}.}{0.5em}{}

\renewenvironment{abstract}
 {\noindent\textbf{\abstractname}\ \ignorespaces}
 {\par\medskip}

\title{\textbf{Predicting passenger injury distributions under uncertainty variables using Gaussian process modeling with GHBMC}}

\author[]{Changmin Baek\thanks{C. Baek(e-mail: min99418@hanyang.ac.kr) is a Master Student of Automotive Engineering, Hanyang University, Seoul, South Korea.}}
\author[]{Junik Cho\thanks{J. Cho(e-mail: junik0322@hanyang.ac.kr) is a Master Student of Automotive Engineering, Hanyang University, Seoul, South Korea.}}
\author[]{Dongjin Lee\thanks{D. Lee(e-mail: dlee46@hanyang.ac.kr) is the corresponding author and an Assistant Professor of Automotive Engineering at Hanyang University, Seoul, South Korea.}}
\affil[]{}

\date{}

\begin{document}

\maketitle

\begin{abstract} 
This work presents a Gaussian Process (GP) modeling method to predict statistical characteristics of injury kinematics responses using Human Body Models (HBM) more accurately and efficiently. We validate the GHBMC model against a 50\%tile male Post-Mortem Human Surrogate (PMHS) test. Using this validated model, we create various postured models and generate injury prediction data across different postures and personalized D-ring heights through parametric crash simulations. We then train the GP using this simulation data, implementing a novel adaptive sampling approach to improve accuracy. The trained GP model demonstrates robustness by achieving target prediction accuracy at points with high uncertainty. The proposed method performs continuous injury prediction for various crash scenarios using just 27 computationally expensive simulation runs. This method can be effectively applied to designing highly reliable occupant restraint systems across diverse crash conditions. 
\end{abstract} 

\section{INTRODUCTION}
\label{intro}
\small 
Human body model (HBM) is an essential tool for developing reliable vehicle restraint systems. HBM overcomes the limitations of the traditional anthropomorphic test devices, such as Hybrid \MakeUppercase{\romannumeral 3}, by offering higher biofidelity that accurately replicates various passenger postures and body types~\cite{widerange}. 
As autonomous driving technology advances, the vehicle control is shifting from human drivers to vehicles, allowing passengers more flexibility in their seating positions. Recognizing the need to evaluate injury mechanics across diverse seating positions, Euro NCAP has announced in Euro NCAP Vision 2030~\cite{ncap} that it will replace conventional regulatory crash tests using ATDs with HBM-based virtual simulations. 
Regulatory crash tests are limited as they evaluate only specific crash scenarios. This makes it challenging to address the full range of real-world accidents, which include various personalized seating conditions---different postures, seat positions, and belt anchor locations. Restraint systems that pass the regulatory crash tests may thus not perform effectively in real-world crash scenarios~\cite{effectiveness,Kasa}.     
To consider various personalized seating environments using HBM, we use Monte Carlo simulations (MCS)~\cite{MC}, which require hundreds or thousands of simulation runs. However, MCS can be computationally burdensome, if not prohibitive, as each HBM simulation takes several hours. We thus use surrogate modeling, such as Polynomial Chaos Expansion~\cite{lee2020practical}, Polynomial Dimensional Decomposition~\cite{rahman2008polynomial,PDD}, Gaussian Process (GP)~\cite{gaussianprocess}, or Reduced Order Modeling~\cite{ROM}, to approximate the relationship between input and output variables, using a limited subset of simulation data to predict outputs for unknown input conditions. Among these surrogate modeling methods, the GP modeling demonstrates excellent ability to predict highly nonlinear injury kinematics responses in the work~\cite{UMTRI,mropti}. However, the GP accuracy could be further improved through better sampling strategies. While the work shows injury distribution, there are many other useful metrics for analyzing injury patterns used in designing restraint systems.   
This study presents a GP modeling method to more accurately predict the statistical characteristics of injury kinematics responses using HBM. First, we validate the Global Human Body Model Consortium (GHBMC) model against a 50th percentile male Post-Mortem Human Surrogate (PMHS) test. Using this validated model, we create various postured models and generate injury prediction data across different postures and personalized D-ring heights through parametric crash simulations. We then train the GP using this data. In the training process, we develop and use a novel adaptive sampling approach to improve GP accuracy. Finally, we use the GP model to generate injury metric distributions, including mean, variance, mode, and value at risk (VaR)~\cite{lee2023multifidelity}. While previous works~\cite{previous1,previous2} examine injury analysis at individual case, we use a statistical approach that quantifies injury risk. This approach enables restraint system designers to predict extreme injury levels in real-world crash scenarios during vehicle development, and thus support creating more reliable restraint systems that better prevent severe injuries. 

\section{METHODS}
\label{body}
\small
\subsection{Baseline Finite Element Model for PMHS Test Conditions}\label{chap:3.1}

\begin{figure}[htbp]
    \centering
    \begin{subfigure}[b]{0.55\textwidth}
        \centering
        \includegraphics[width=\textwidth]{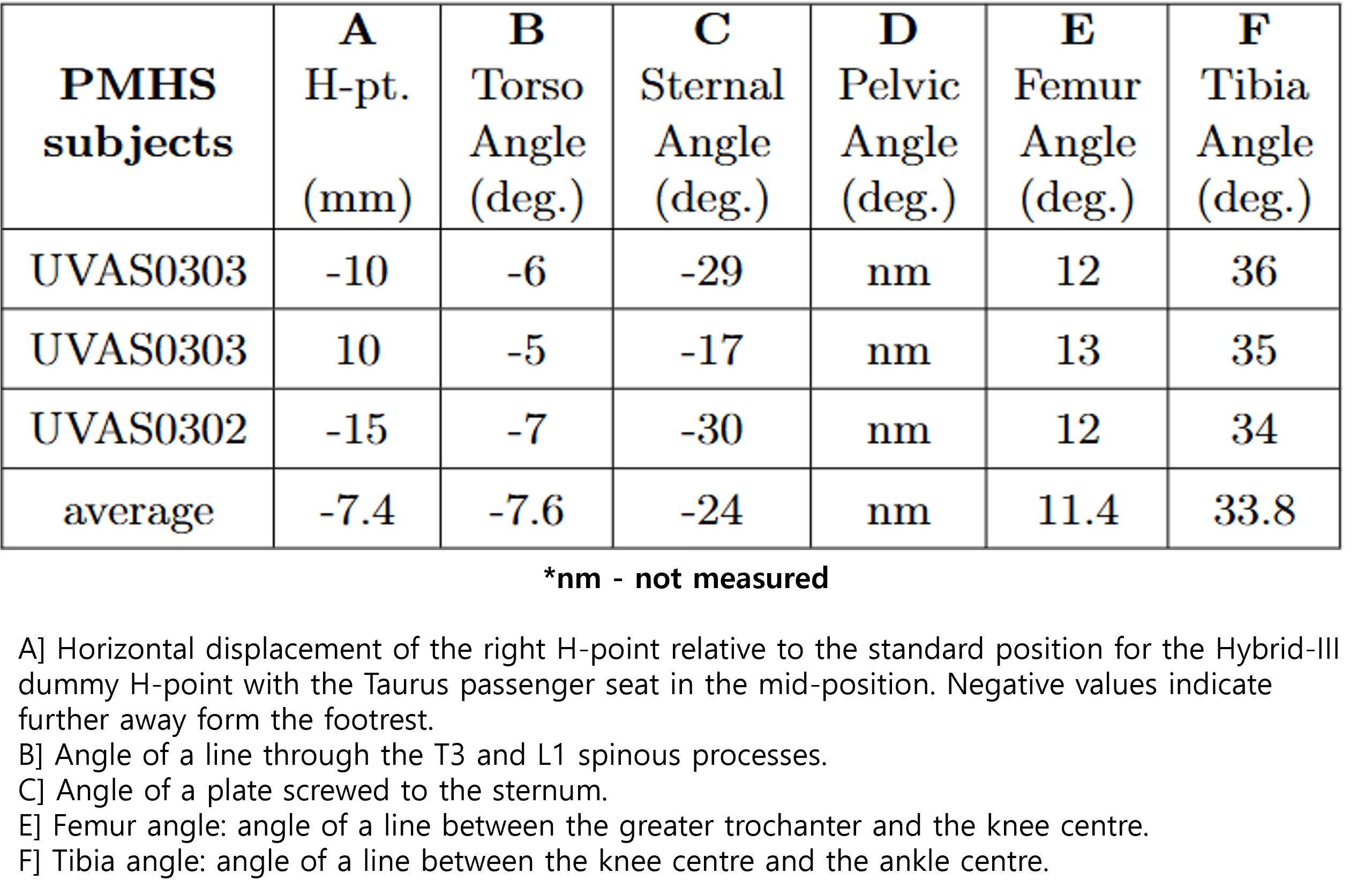}
        \caption{PMHS subject positioning table and descriptions}
        \label{fig:pmhs_table}
    \end{subfigure}
    \hfill
    \begin{subfigure}[b]{0.4\textwidth}
        \centering
        \includegraphics[width=0.9\textwidth]{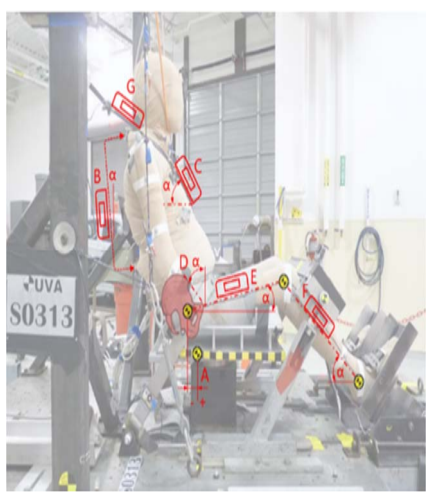}
        \caption{PMHS positioning Standard with angle definitions}
        \label{fig:pmhs_image}
    \end{subfigure}
    \caption{PMHS subject posture measurements and visualized positioning standard~\cite{PMHStest,PMHS2015}}
    \label{fig:pmhs_subplot}
\end{figure}

\begin{figure}[H]
    \centering
    \begin{minipage}[b]{0.30\textwidth}
        \begin{subfigure}[b]{\textwidth}
            \centering
            \includegraphics[width=\linewidth]{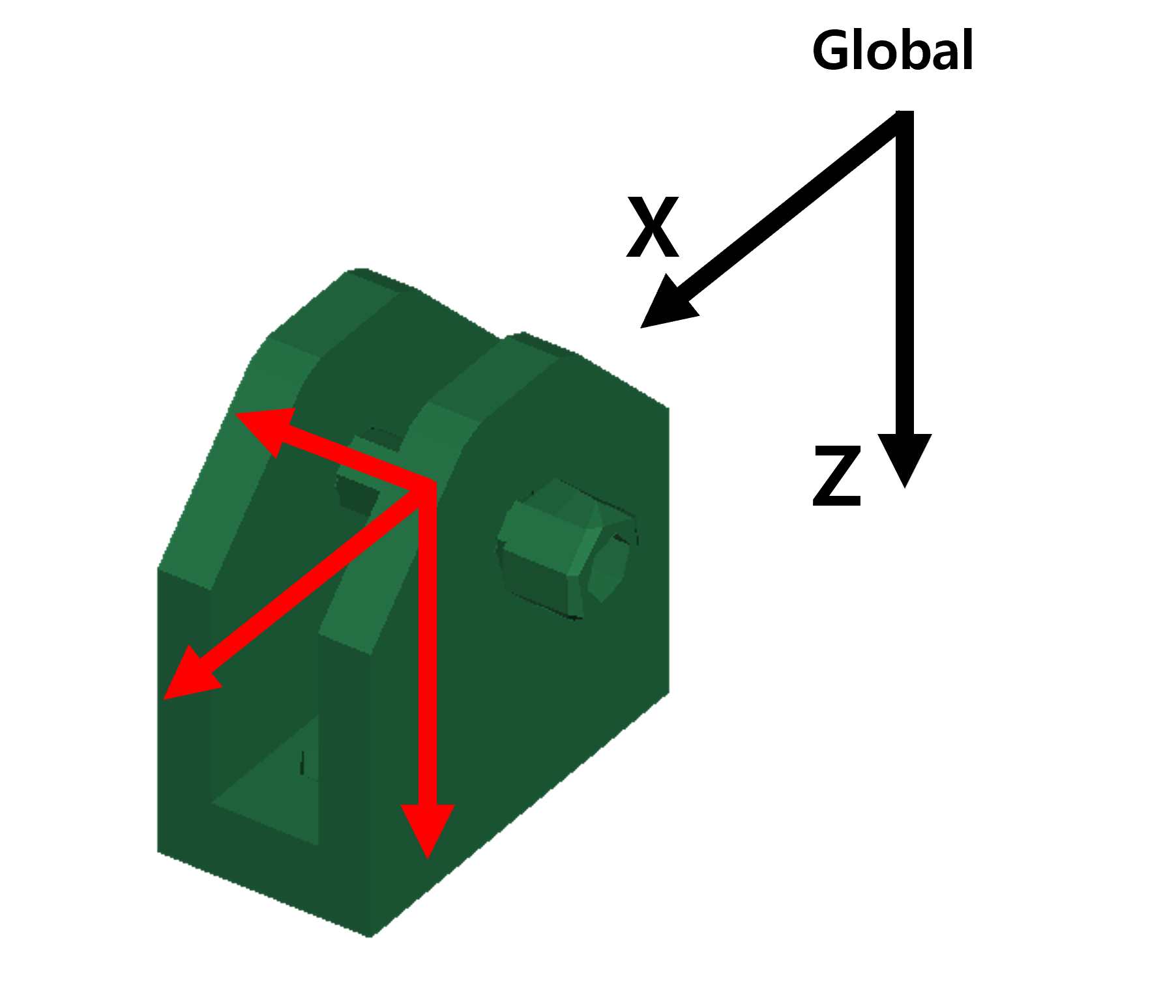}
            \caption{Buck coordinate system}
            \label{fig:buck_cs}
        \end{subfigure}

        \vspace{1em}

        \begin{subfigure}[b]{\textwidth}
            \centering
            \includegraphics[width=\linewidth]{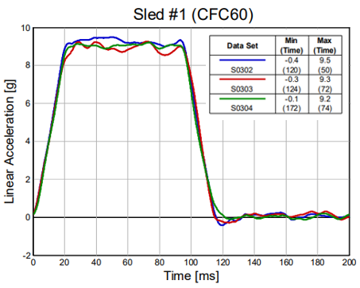}
            \caption{9g sled impact pulse}
            \label{fig:sled_pulse}
        \end{subfigure}
    \end{minipage}
    \hfill
    \begin{minipage}[b]{0.6\textwidth}
        \begin{subfigure}[b]{\textwidth}
            \centering
            \includegraphics[width=\linewidth]{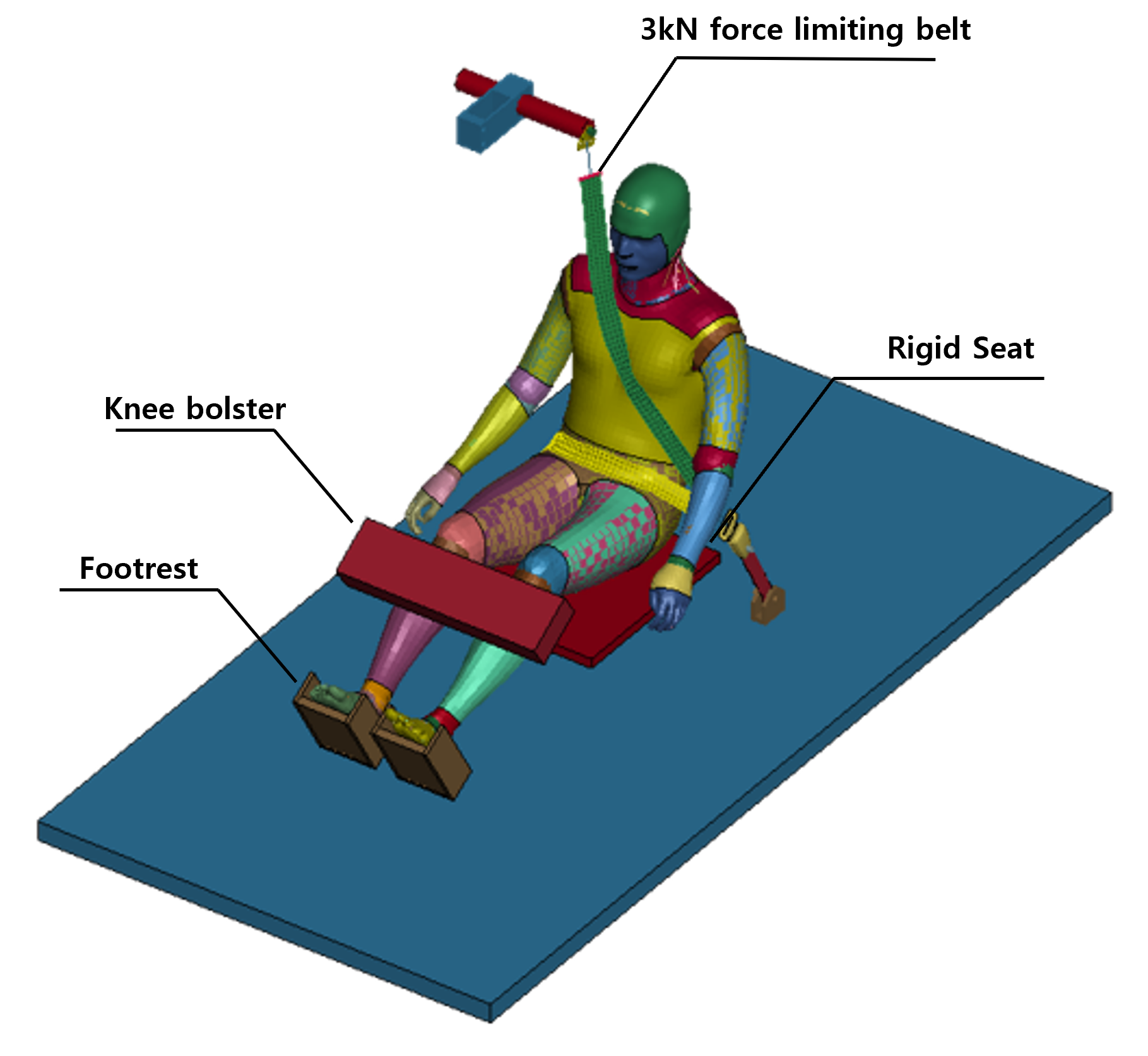}
            \caption{FE model setup}
            \label{fig:fe_model}
        \end{subfigure}
    \end{minipage}

    \caption{Overview of simulation setup: (a) Buck coordinate system, (b) Sled impact pulse, (c) FE model setup}
    \label{fig:simulation_setup}
\end{figure}

We construct a baseline finite element model based on experimental results~\cite{PMHStest,PMHS2015} from frontal impact tests conducted at University of Virginia using three male PMHS subjects (weight = 68 ± 2 kg, height = 1,780 ± 50 mm). We use the 50 \%tile HBM model (GHBMC M50-OS v2-3), initially positioning it according to specific criteria derived from the PMHS test conditions. Figure~\ref{fig:pmhs_table} presents these criteria, including torso angle, femur angle, and tibia angle. 
To do so, we use LS-DYNA's built-in functionality~\cite{prepositioing} based on the averaged positional data from the three PMHS subjects and then use gravity settling to achieve a realistic occupant posture. Figure~\ref{fig:simulation_setup} presents the simulation model for the experimental sled setup consisted of a 3-kN force-limiting 3-point belt and rigid components (knee bolster, seat plate, and footrest). We finally implement the crash sled simulation with the 9g impact pulse averaged in the three PMHS tests, as shown in  Figure~\ref{fig:sled_pulse}. We validate the model using PHMS test data, comparing trajectory and maximum excursion values of Head, T1, L1, and force data from the upper shoulder belt and $Z$-axis seat. We measure these kinematic results in the buck coordinate system (referenced to the lap belt anchor), shown in Figure~\ref{fig:buck_cs}, and calculate acceleration values using the SAE coordinate system.

\subsection{Parametric models for various occupant postures and D-ring heights}
Using the baseline model, we generate a parametric model with two design variables: torso angle and D-ring $Z$ position. Each variable varies across five levels within its range (torso angles from -10 $^\circ$ to 10 $^\circ$ and D-ring $Z$ position from -50 mm to 50 mm, relative to the baseline model), creating 25 different occupant configurations. From these parametric model simulations, we compute the Head Injury Criterion (HIC 15) which is an index that evaluates the likelihood of head injury by measuring head acceleration over a 15-millisecond interval and maximum T1 $X$-axis acceleration ($a_{\text{T1,max}}$) values. 
\begin{problem}\label{prob1}
Using several dozen simulation datasets, we accurately and efficiently predict crash passenger injury distributions and their statistical properties, including mean, variance, and value at risk. 
\end{problem}
\subsection{The Gaussian Process model}
We use the Gaussian Process (GP) model to predict the injury responses or matrices, such as (HIC~15) and maximum T1 $X$-axis acceleration ($a_{\text{T1,max}}$), for unknown inputs $\mathbf{x}=(x_1,x_2)^{\intercal}\in\mathbb{R}^2$, $x_1$=torso angle ($^circ$) and $x_2$=D-ring $Z$ position (mm). Using several dozen ($L=25$) datasets obtained from simulation runs, we train the GP to predict the distribution or associated statistical properties (mean, variance, and value at risk) of the injury matrices.

For given input-output datasets $\{\mathbf{x}^{(l)}, f_{\mathrm{HIC15}}(\mathbf{x}^{(l)}) \}_{l=1}^L$ and $\{\mathbf{x}^{(l)}, f_{a_{\text{T1,max}}}(\mathbf{x}^{(l)}) \}_{l=1}^L$, we can approximate output function HIC 15 $f_{\text{HIC15}}(\cdot)$ and Max T1 $X$-acceleration $f_{a_{\text{T1,max}}}(\cdot)$ using a Gaussian Process model as follows:
\begin{align}
    f_{\text{HIC15}}(\mathbf{x}) &\sim \mathcal{GP} \left( \mu_{\text{HIC15}}(\mathbf{x}), k_{\text{HIC15}}(\mathbf{x}, \mathbf{x'}) \right), \label{eq:hic15} \\
    f_{a_{\text{T1,max}}}(\mathbf{x}) &\sim \mathcal{GP} \left( \mu_{a_{\text{T1,max}}}(\mathbf{x}), k_{a_{\text{T1,max}}}(\mathbf{x}, \mathbf{x'}) \right). \label{eq:maxt1acc}
\end{align}
Here, for input $\mathbf{x}$, $\mu_{\text{HIC15}}(\mathbf{x})$ and $\mu_{a_{\text{T1,max}}}(\mathbf{x})$ are the mean for injury matrices, HIC15 and $a_{\text{T1,max}}$, respectively. For input $\mathbf{x}$ and a given training input data $\mathbf{x}'$, $k_{\text{HIC15}}(\mathbf{x}, \mathbf{x'})$ and $k_{a_{\text{T1,max}}}(\mathbf{x}, \mathbf{x'})$ are the covariances for injury matrices $\mathbf{x}$, HIC15 and $a_{\text{T1,max}}$, respectively. In this study, we set the mean function to \emph{zero} and adapt the Matérn kernel as the covariance function, which provides flexibility in modeling complex relationships within the input data.
For a new input vector $\mathbf{x^*}$, the GP model predicts the corresponding injury response by computing the posterior distribution based on the training data.
\begin{equation}
    \tilde{y}(\mathbf{x^*}) \sim \mathcal{N} \left( \tilde{\mu}(\mathbf{x^*}), \tilde{\sigma}^{2}(\mathbf{x^*}) \right).
    \label{eq:gp_prediction}
\end{equation}
The predictive mean \( \tilde{\mu}(\mathbf{x^*}) \) represents the expected value of the injury response, while the predictive variance \( \tilde{\sigma}^{2}(\mathbf{x^*}) \) quantifies the uncertainty of the prediction. These are computed as follows:
\begin{gather}
    \tilde{\mu}(\mathbf{x^*}) = \mathbf{K}_*^T (\mathbf{K} + \sigma^2 \mathbf{I})^{-1} \mathbf{y}, \label{eq:mu} \\
    \tilde{\sigma}^{2}(\mathbf{x^*}) = k(\mathbf{x^*}, \mathbf{x^*}) - \mathbf{K}_*^T (\mathbf{K} + \sigma^2 \mathbf{I})^{-1} \mathbf{K}_*. \label{eq:sigma}
\end{gather}
Here, $\mathbf{K}$ is the covariance matrix computed from the training data, $\mathbf{K}_*$ represents the covariance between the training data and the test input \( \mathbf{x^*} \), and \( \sigma^2 \) is the noise variance. This formulation allows the GP model to provide a point prediction for the injury response and an uncertainty measure, which is particularly useful for assessing the reliability of predictions in injury risk analysis.  
\subsection{Improving the Predictive Accuracy of the GP Model}\label{chap:3.4}
\begin{figure}[H]
    \centering
    \includegraphics[width=0.8\textwidth]{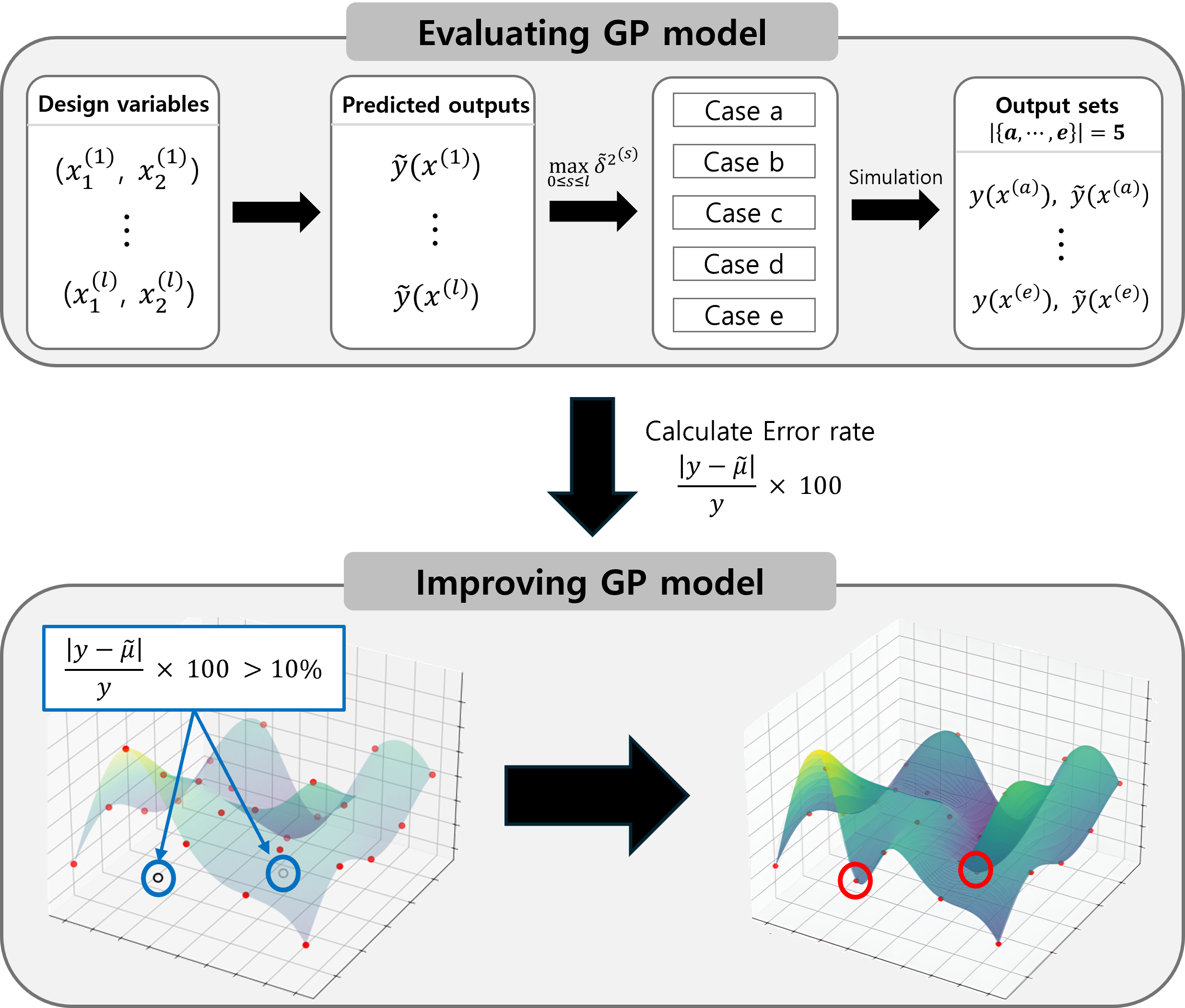}
    \caption{Improving the accuracy of the GP model through adaptive training}
    \label{fig:active_learning}
\end{figure}

Figure~\ref{fig:active_learning} presents the progress of evaluating and improving the accuracy of the GP model initially trained using a dataset of 25 simulations. We first identify the predictive output data where the GP model exhibits the highest variance $\tilde{\sigma}^2(\mathbf{x}^*)$. These data points are considered to have the lowest predictive accuracy. Using these dataset as a test dataset, we generate five additional datasets via simulation. If the predictive accuracy for these five output data shows less than 10\% error between GP-based predictive data and simulation data, we include these simulation datasets into the training set to ensure that the worst case predictive error decreases to less than 10\%.  

\subsection{Predicting injury distributions}
We aim to predict the distribution of outputs for injury matrices HIC15 and maximum T1 $X$-acceleration, as presented in Problem~\ref{prob1}. Using the trained GP model from Section~\ref{chap:3.4}, we can efficiently and accurately generate tens of thousands of datasets to predict the distribution. Readers need to note that obtaining these datasets through simulation runs is computationally extensive.

For the input vector $\mathbf{x}=(x_1,x_2)^{\intercal}$, $x_1$=torso angle and $x_2$=D-ring $Z$ position, we generate input data $\{\mathbf{x}^{(l)}\}_{l=1}^{10,000}$, following a uniform distribution within $[-10 ^\circ,10 ^\circ]$ for the torso angle and $[-50\text{mm},50\text{mm}]$ for the D-ring $Z$ position via Latin Hypercube Sampling~\cite{LHS,LHS2}. We then use the trained GP model's eq.\eqref{eq:mu} to predict the corresponding outputs $\{\tilde{\mu}{\text{HIC15}}(\mathbf{x}^{(l)}),\tilde{\mu}{a_{\text{T1,max}}}(\mathbf{x}^{(l)})\}_{l=1}^{10,000}$. Finally, we estimate the distribution using the empirical probability density function, providing stochastic properties such as the mean, variance, and value at risk (VaR) at 90\%tile and 99\%tile. 

\begin{figure}[H]
    \centering
    \includegraphics[width=0.7\textwidth]{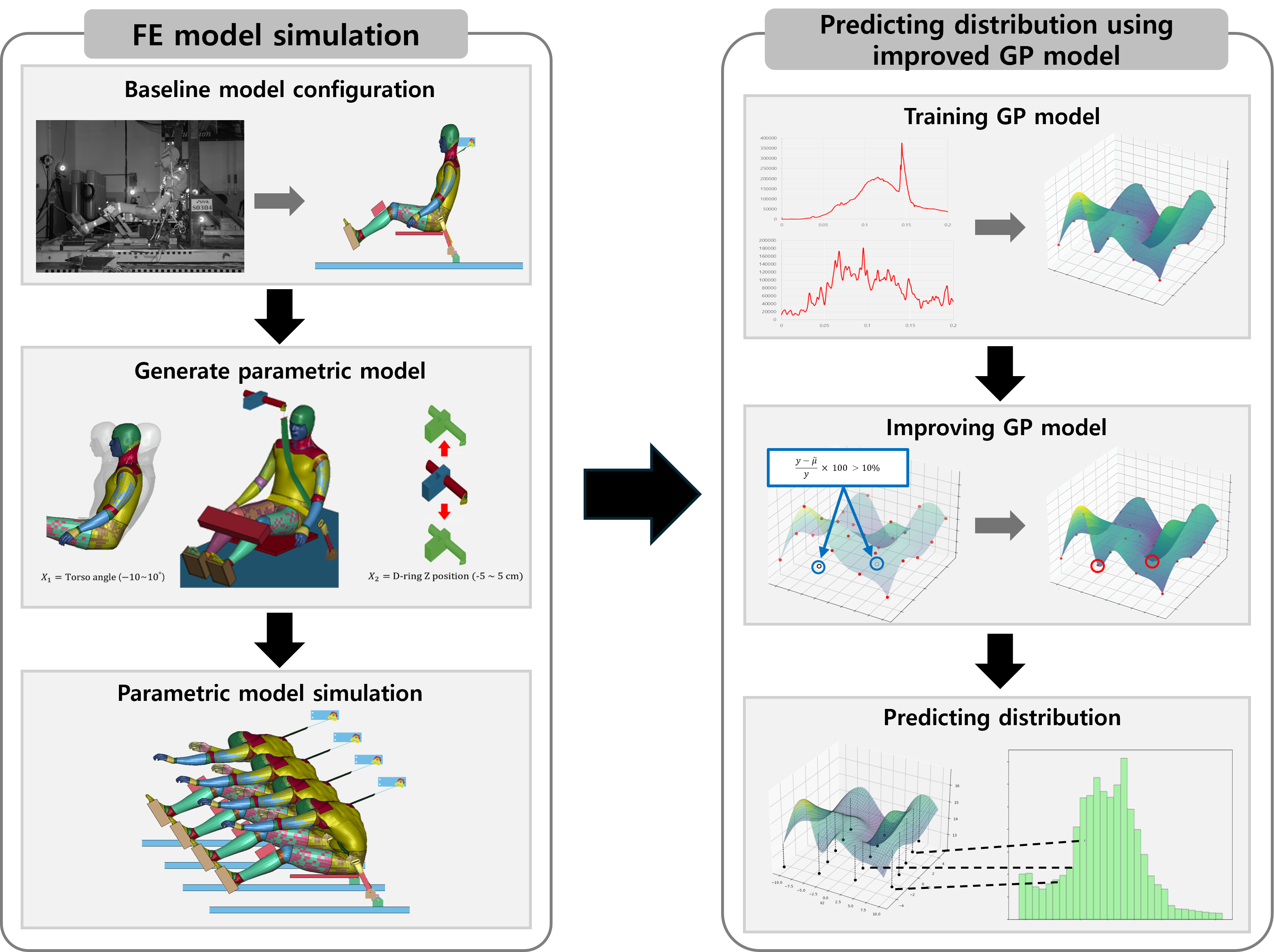}
    \caption{Framework of the surrogate-based injury prediction method}
    \label{fig:framework}
\end{figure}

Figure~\ref{fig:framework} presents the overall framework for predicting injury metrics distribution. The framework utilizes HBM-based simulation data to train the GP models, as explained in Sections~\ref{chap:3.1}--\ref{chap:3.4}.

\begin{figure}[H]
    \centering
    \begin{minipage}{0.45\textwidth} 
        \centering
        \includegraphics[width=\textwidth]{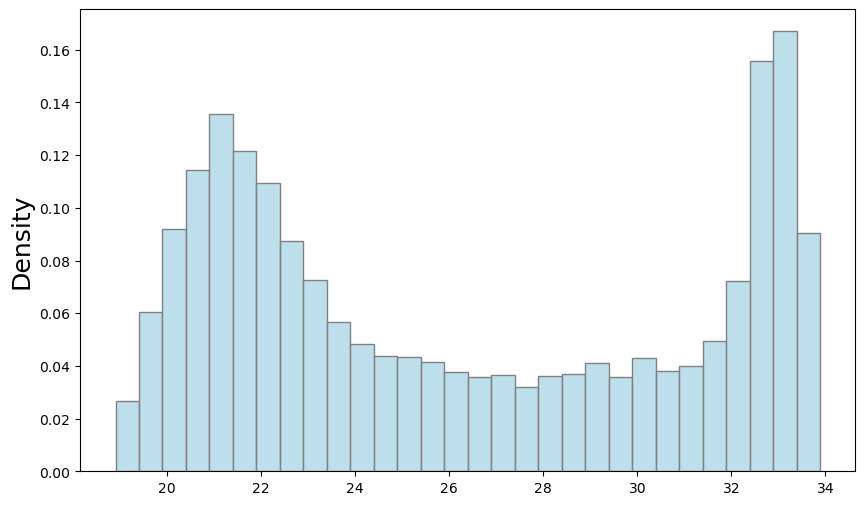}
    \end{minipage}
    \hfill
    \begin{minipage}{0.45\textwidth} 
        \centering
        \includegraphics[width=\textwidth]{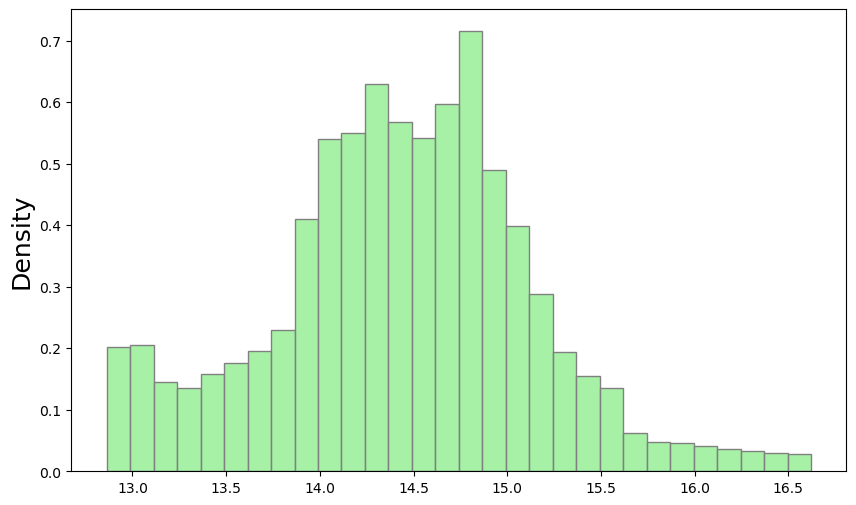}
    \end{minipage}    
        \caption{Predicted distributions of HIC15 (left) and $a_{\text{T1, max}}$ (right)}
        \label{fig:distribution}
\end{figure}

\section{RESULTS}
\label{Conclusion}
\small
\subsection{baseline model validation}
We validate the baseline simulation model by comparing trajectories and peak excursions of the HEAD, T1, and L2 in the X and Z directions, along with upper shoulder belt force and $Z$-axis seat force values against the PMHS test results, as discussed in Section~\ref{chap:3.1}.

\begin{figure}[H]
  \centering
  \begin{minipage}{0.32\textwidth}
    \includegraphics[width=\linewidth]{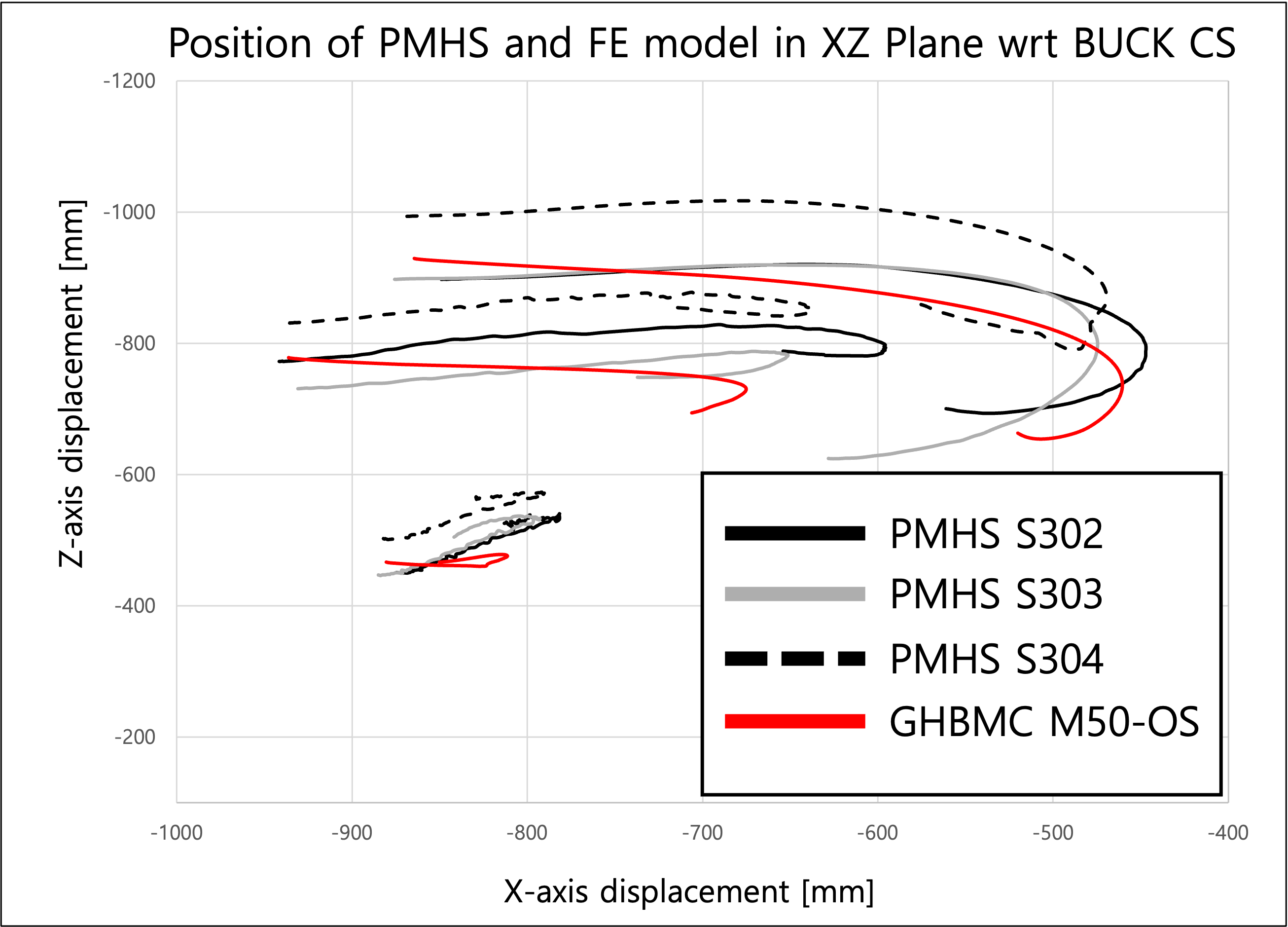}
  \end{minipage}
  \begin{minipage}{0.32\textwidth}
    \includegraphics[width=\linewidth]{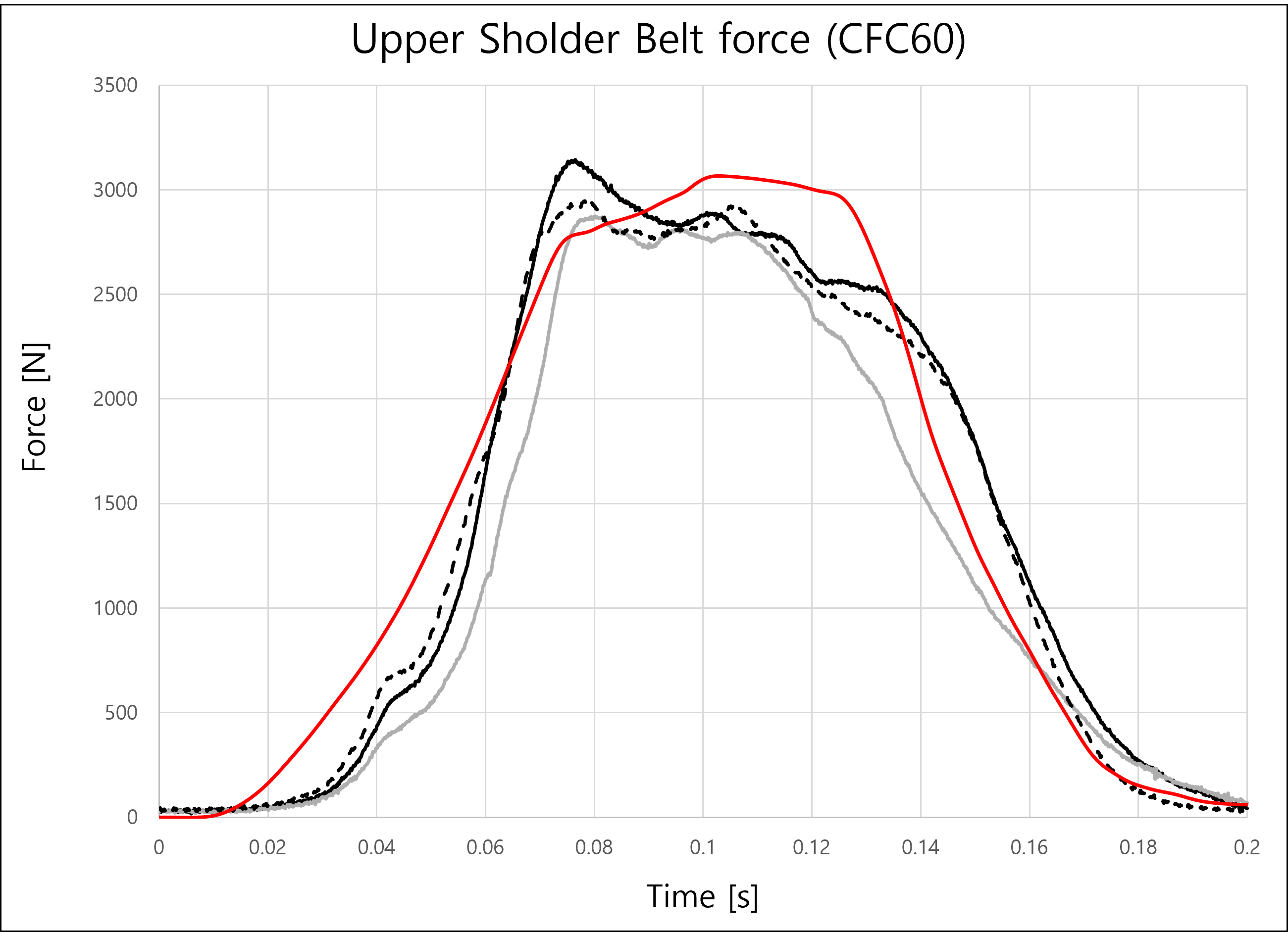}
  \end{minipage}
  \begin{minipage}{0.32\textwidth}
    \includegraphics[width=\linewidth]{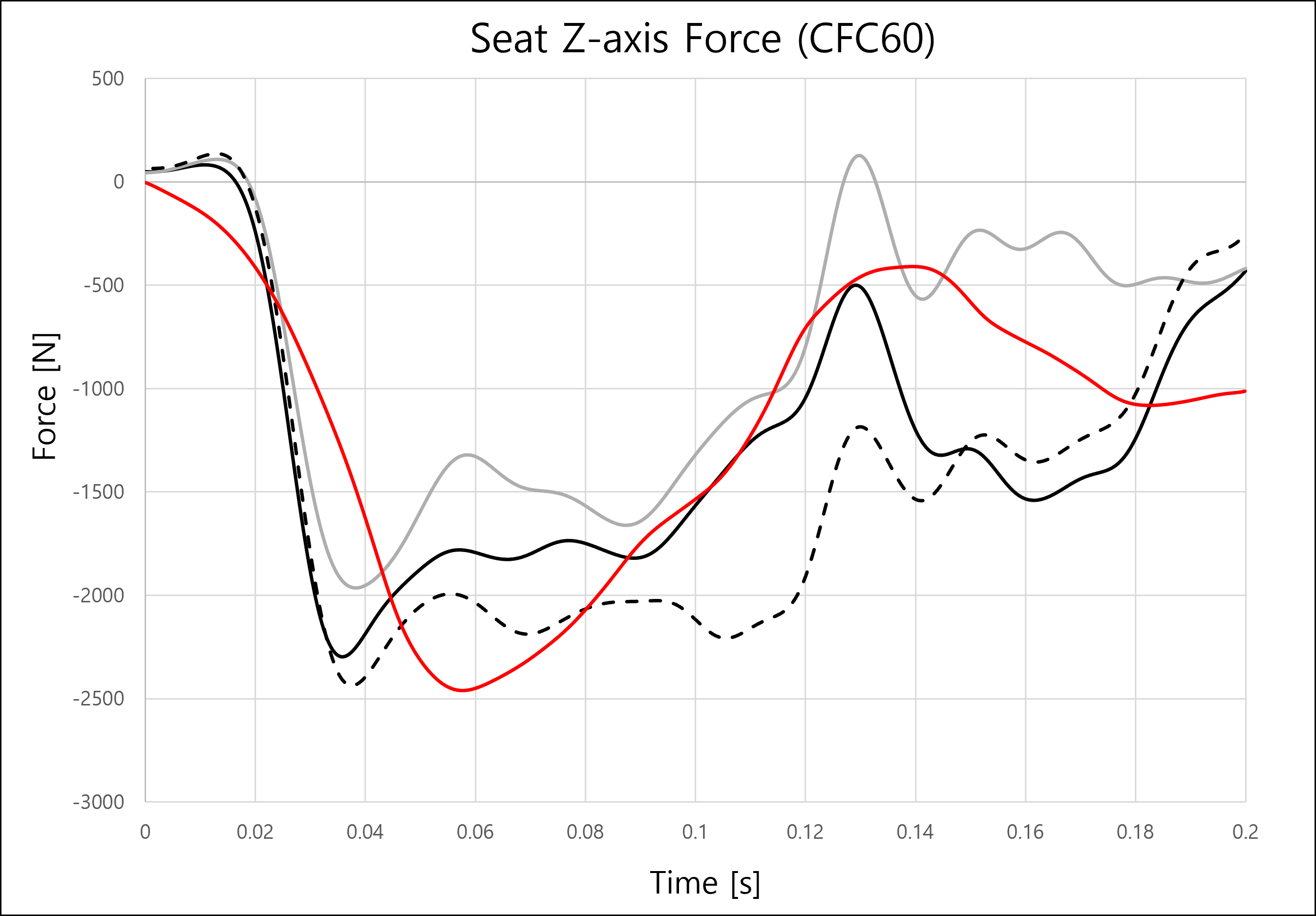}
  \end{minipage}
  \caption{Results of FE model validation with PMHS tests}
  \label{fig:model_validation_results}
\end{figure}

Figure~\ref{fig:model_validation_results} shows that the model's kinematics follow a similar trend to the PMHS tests. The HEAD's peak excursion values present a 0.3\% deviation in the X direction and a 13.2\% deviation in the Z direction. 
For T1, the excursion in the X direction has a 5.19\% error. 
The maximum belt force differs by 7\% between the test and simulation results, and the minimum $Z$-direction seat force shows a similar 7\% deviation.
Table~\ref{tab:comparison} summarizes the maximum displacement values for the HEAD, T1, and L2 markers in both $X$ and $Z$ directions, comparing the baseline FE model simulation with PMHS test results. The predicted displacement values of injury kinematics closely match the PMHS results, showing 0.2-17\% error with the exception of $Z$-direction excursion. 

\vspace{-10pt} 

\sisetup{
	detect-weight = true,
	detect-inline-weight = math,
	output-decimal-marker = {.},
	table-align-text-post = false,
	table-number-alignment = right,
	round-mode = places,
	drop-zero-decimal = true
}

\begin{table}[H]
	\centering
	\renewcommand{\arraystretch}{0.9}
	\captionsetup{skip=5pt}
	\begin{tabular}{
			>{\centering\arraybackslash}p{3.8cm}
			S[table-format=4.0, table-number-alignment=right, table-alignment=center]
			S[table-format=4.1, table-number-alignment=right, table-alignment=center, round-precision=1]
			S[table-format=4.2, table-number-alignment=right, table-alignment=center, drop-zero-decimal=false]
		}
		\toprule
		\textbf{Variable} &
		{\textbf{PMHS avg. (mm)}} &
		{\textbf{Simulation (mm)}} &
		{\textbf{Error (\%)}} \\
		\midrule
		Head X Excursion & 401  & 400     & 0.25 \\
		Head Z Excursion & 250  & 283     & 13.20 \\
		\hspace*{0.6em}T1\hspace*{0.6em}X Excursion   & 308  & 292.2   & 5.13 \\
		\hspace*{0.6em}T1\hspace*{0.6em}Z Excursion   & -53  & 74      & 239.62 \\
		\hspace*{0.6em}L2\hspace*{0.6em}X Excursion   & 92   & 76      & 17.39 \\
		\hspace*{0.6em}T1\hspace*{0.6em}Z Excursion   & -84  & -21     & 75.00 \\
		\bottomrule
	\end{tabular}
	\caption{Peak excursion differences between simulation results and PMHS average values across body regions}
	\label{tab:comparison}
\end{table}

\subsection{Improving the Gaussian Process model}
\begin{figure}[H]
    \centering

    \begin{subfigure}[b]{0.48\textwidth}
        \centering
        \includegraphics[width=\textwidth]{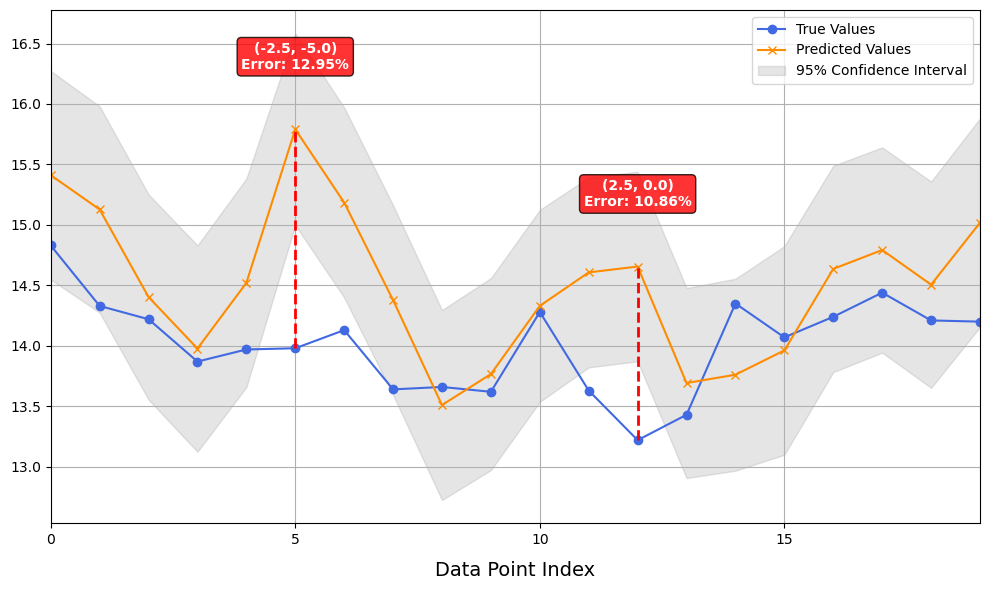}
        \caption{GP initially trained with 25 datasets}
        \label{fig:t1acc_initial}
    \end{subfigure}
    \hfill
    \begin{subfigure}[b]{0.48\textwidth}
        \centering
        \includegraphics[width=\textwidth]{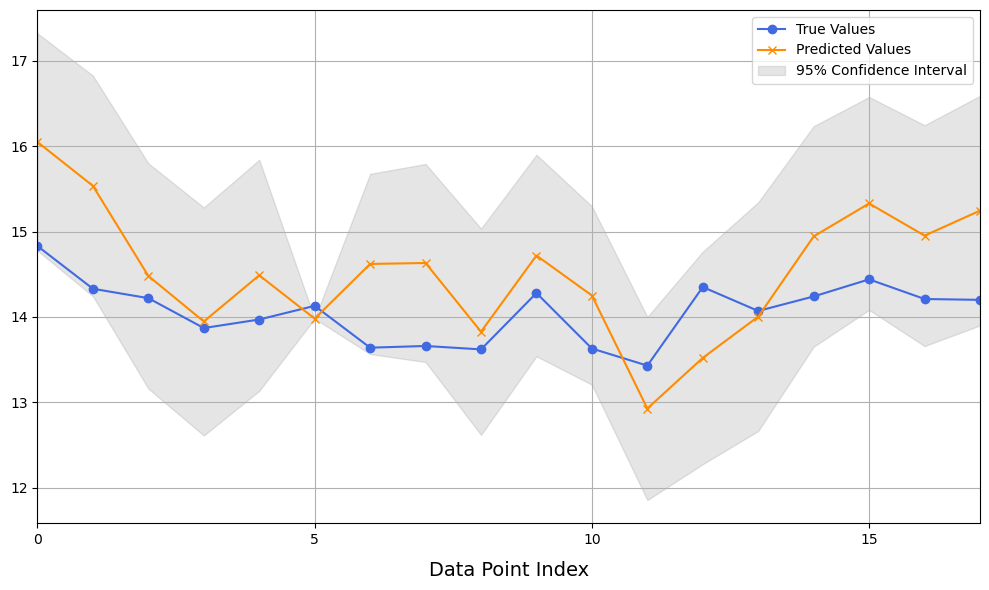}
        \caption{GP adaptively trained with 27 datasets}
        \label{fig:t1acc_enhanced}
    \end{subfigure}

    \caption{Comparison of $a_{\text{T1,max}}$ predictions between GP models using 25 and 27 datasets}
    \label{fig:t1acc_compare}
\end{figure}

We evaluate the accuracy of the GP model, which is initially trained on 25 datasets, by comparing its predictions to the simulation results. For the HIC15 prediction model, the prediction errors across all 5 test datasets remain within the acceptable threshold of 10\% relative to the simulation results. 
However, for the Max T1 $X$ acceleration ($a_{\text{T1,max}}$), the GP model exceeded the acceptable prediction error threshold of 10\%, showing percentage errors of 12.95\% and 10.86\% at two input points (-2.5 $^\circ$, -5 mm) and (2.5 $^\circ$, 0 mm), respectively. To achieve the target accuracy, we incorporated these two points into the training dataset for the $a_{\text{T1,max}}$ model, resulting in a total of 27 training points. 
We generate five test simulation data at the input points with the high GP predictive variance $\tilde{\sigma}^2(\mathbf{x})$ within the input domain, which represents the location where the GP model's prediction has the great uncertainty. Figure~\ref{fig:t1acc_compare} presents the detailed prediction results for two GP models---one (left) initially trained with 25 simulation datasets and another (right) adaptively trained with 27 dataset for maximum T1 $X$-acceleration ($a_{\text{T1,max}}$). The error for the test data compared to the simulation results at the test input points satisfies the target criterion. 

\subsection{Statistics of predicted distributions}
From the predicted distributions of HIC15 and Maximum T1-$X$ acceleration ($a_{\text{T1,max}}$), we calculate statistical quantities---mean, standard deviation, mode, maximum, minimum values, and Value at Risk at 90\%tile and 95\%tile---to quantitatively analyze passenger injury under various crash conditions.

Table~\ref{tab:statistics_var} presents a summary of the estimated statistical quantities for the HIC15 and Maximum T1 $X$ acceleration. For the HIC15 distribution, the mean is 26.24 with a standard deviation of 4.88. This indicates that HIC15 values under the defined crash conditions are likely to fall between 21.36 and 31.12. The mode is 32.86, indicating the most frequent injury risk level. The maximum value reaches 33.59—7.35 above the mean—representing the highest expected HIC15 value under these crash conditions. The Value at Risk analysis reveals a VaR at 90\%tile of 32.8 and a VaR at 90\%tile of 33.09. 

For the Maximum T1 $X$ acceleration ($a_{\text{T1,max}}$), the mean value is 14.41 $\mathrm{m/s^2}$ with a standard deviation of 0.7 $\mathrm{m/s^2}$, indicating that values are highly likely to fall within the range of 13.71 $\mathrm{m/s^2}$–15.11 $\mathrm{m/s^2}$ under the crash conditions. The mode of the distribution is 14.8 $\mathrm{m/s^2}$. The maximum value reaches 16.64 $\mathrm{m/s^2}$, which is 2.23 $\mathrm{m/s^2}$ higher than the mean, representing the highest expected T1 acceleration under the given crash conditions. Value at Risk (VaR) analysis shows that the VaR at 90\%tile is 15.22 $\mathrm{m/s^2}$ and the VaR at 95\%tile is 15.5 $\mathrm{m/s^2}$.

\begin{figure}[H] 
    \centering
    \begin{subfigure}{0.49\textwidth}
        \centering
        \includegraphics[width=\linewidth]{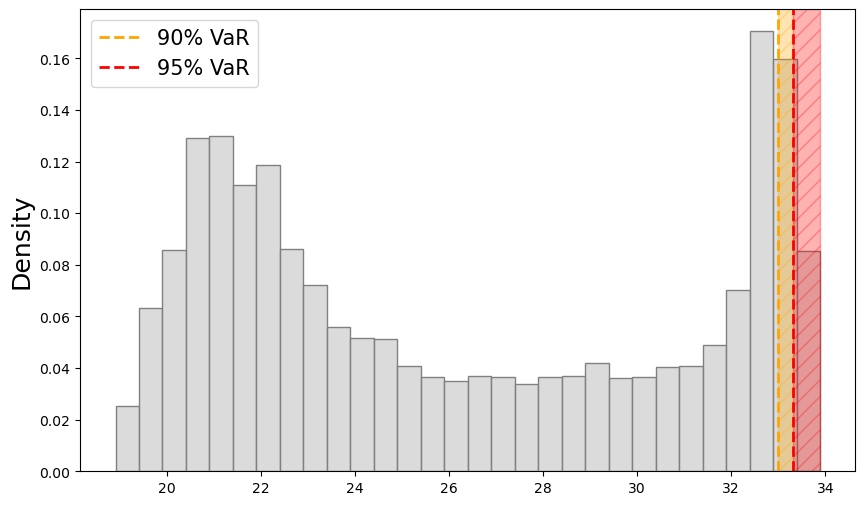}
        \caption{HIC15}
        \label{fig:var_hic15}
    \end{subfigure}
    \begin{subfigure}{0.49\textwidth}
        \centering
        \includegraphics[width=\linewidth]{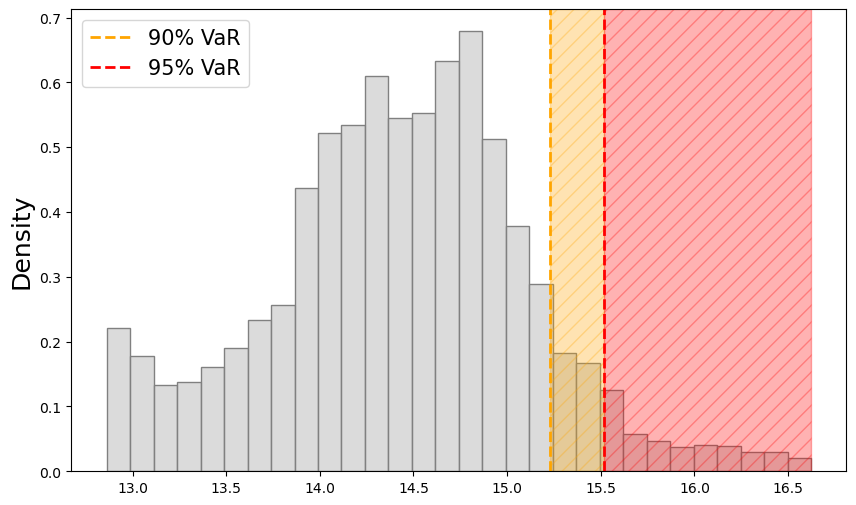}
        \caption{Maximum T1 $X$ acceleration ($a_{\text{T1,max}}$)}
        \label{fig:var_t1acc}
    \end{subfigure}
    \caption{Value at Risk (VaR) for injury metrics distribution}
    \label{fig:var_comparison}
\end{figure}

Figure~\ref{fig:var_comparison} shows the distributions of the two injury metrics.  The shaded areas indicate regions that exceed the 90\% and 95\% VaR thresholds.
\begin{table}[H]
    \centering
    \renewcommand{\arraystretch}{1.2} 
    \begin{tabular}{lccccccc}
        \toprule
        & \textbf{Mean} & \textbf{Std} & \textbf{Mode} & \textbf{Min} & \textbf{Max} & \textbf{VaR (90\%tile)} & \textbf{VaR (95\%tile)} \\ 
        \midrule
        HIC15  & 26.24 & 4.88 & 32.86 & 18.95 & 33.59 & 32.8  & 33.09 \\ 
        $a_{\text{T1,max}}$  & 14.41 & 0.7  & 14.8  & 12.83 & 16.64 & 15.22 & 15.5  \\ 
        \bottomrule
    \end{tabular}
    \caption{Statistical properties of injury metrics distributions}
    \label{tab:statistics_var}
\end{table}
\section{DISCUSSION}

This study introduces a Gaussian Process surrogate modeling approach that utilizes HBM simulation data. The method enables both prediction and statistical analysis of injury distributions across diverse driving scenarios. By generating continuous rather than discrete injury distributions, this approach allows for more comprehensive statistical analysis compared to previous studies~\cite{previous1,previous2}.
The analysis of the GHBMC-based model revealed discrepancies exceeding 13\% in the kinematic behavior of the head and T1 in the Z direction, as well as L2 in both X and Z directions, when compared to experimental results. Furthermore, while the PMHS experiment demonstrated upward kinematic motion for T1 and L2 from their initial positions, the GHBMC simplified model showed a continuous downward trend. This difference likely stems from the simplified model's flesh components (skin, muscle, and adipose tissues) not being based on actual human tissue data. Other studies using the GHBMC simplified model under similar experimental conditions have reported comparable findings.
The proposed method for GP modeling allows designers to quantitatively assess injury mitigation performance through statistical indicators like variance, leading to more effective safety design.
\section{CONCLUSION}
This work proposes a Gaussian Process (GP) modeling method to predict statistical characteristics of injury kinematics responses using Human Body Models (HBM) more accurately and efficiently. We first validated the GHBMC model against a 50\%tile male Post-Mortem Human Surrogate (PMHS) test. Using this validated model, we created various postured models and generated injury prediction data across different postures and personalized D-ring heights through parametric crash simulations. We then trained the GP using this simulation data, implementing a novel adaptive sampling approach to improve accuracy. The trained GP model demonstrated robustness by achieving target prediction accuracy at points with high uncertainty. The proposed method performed continuous injury prediction for various crash scenarios using just 27 computationally expensive simulation runs. This method can be effectively applied to designing highly reliable occupant restraint systems across diverse crash conditions.

\section*{ACKNOWLEDGMENT}
This research was supported by Korea Institute for Advancement of Technology(KIAT) grant funded by the Korea Government(MOTIE) (P0017120, The Competency Development Program for Industry Specialist). We thank Sung Rae Kim from Hyundai Motor Company for helpful discussion about the GHBMC model. 
\bibliographystyle{unsrt}
\bibliography{paper_ircobi_Baek}

\section*{APPENDIX}
\appendix
\section*{Kinematic comparision between PMHS with GHBMC}

\begin{figure}[H]
    \centering
    \includegraphics[width=1.0\textwidth]{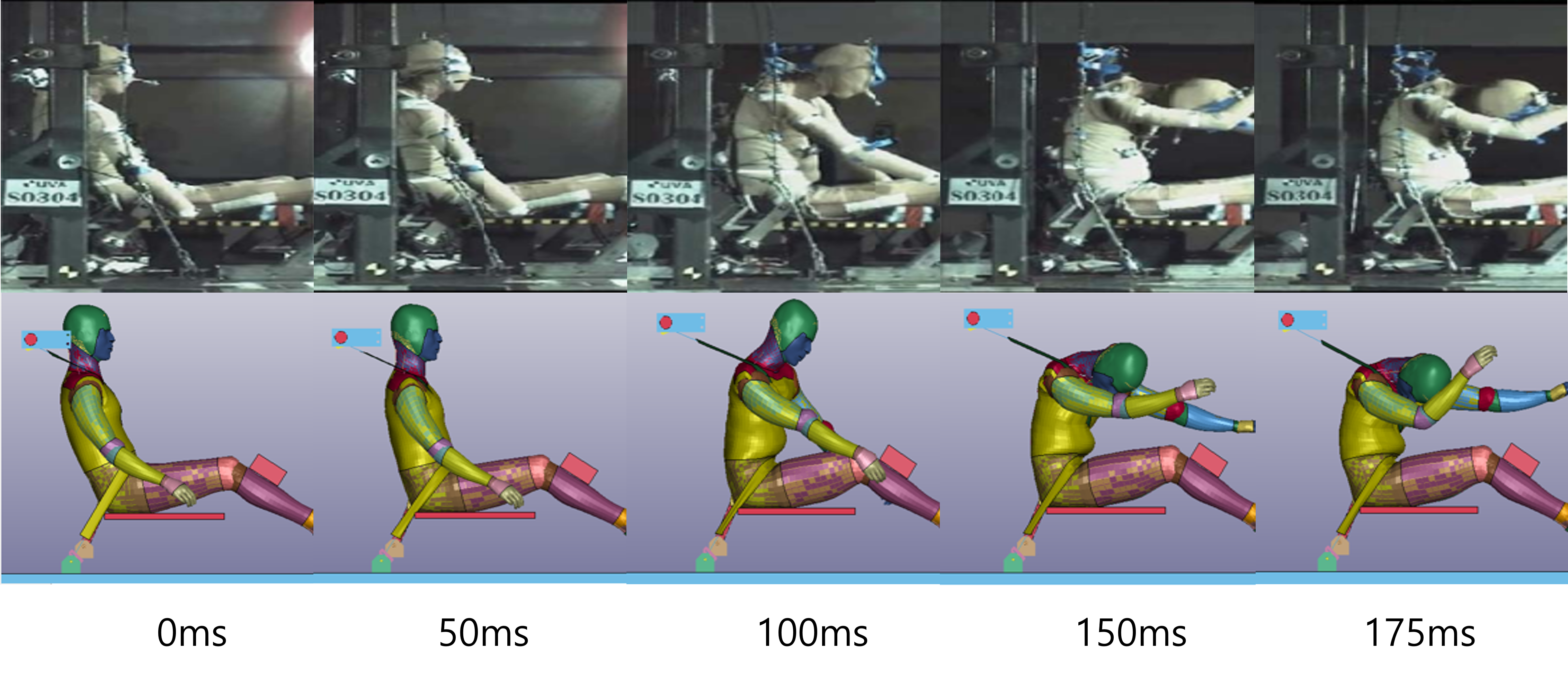}
    \label{fig:appendixA}
\end{figure}
\section*{Parametric model simulation results table}
\begin{table}[htbp]
    \centering
    \renewcommand{\arraystretch}{1.0}  
    \begin{tabular}{|c|c|c|c|c|}
        \hline
        Case number & Torso Angle ($x_1$) & D-ring $Z$ Position ($x_2$) & HIC15 & $a_{\text{T1,max}}$~[$\mathrm{m/s^2}$] \\
        \hline
        1  & \multirow{5}{*}{-10} & -5   & 20.46  & 13.74 \\
        2  &  & -2.5 & 19.44  & 14.32 \\
        3  &  & 0    & 18.91  & 13.68 \\
        4  &  & 2.5  & 19.44  & 13.33 \\
        5  &  & 5    & 19.34  & 13.64 \\
        \hline
        6  & \multirow{5}{*}{-5} & -5   & 21.77  & 16.33 \\
        7  &  & -2.5 & 21.93  & 15.29 \\
        8  &  & 0    & 21.38  & 14.61 \\
        9  &  & 2.5  & 21.42  & 13.92 \\
        10 &  & 5    & 22.00  & 14.71 \\
        \hline
        11 & \multirow{5}{*}{0}  & -5   & 26.41  & 14.82 \\
        12 &  & -2.5 & 25.02  & 14.86 \\
        13 &  & 0    & 25.84  & 14.35 \\
        14 &  & 2.5  & 25.11  & 13.20 \\
        15 &  & 5    & 23.53  & 13.08 \\
        \hline
        16 & \multirow{5}{*}{5} & -5   & 32.00  & 14.16 \\
        17 &  & -2.5 & 32.91  & 14.46 \\
        18 &  & 0    & 31.20  & 15.23 \\
        19 &  & 2.5  & 31.23  & 14.21 \\
        20 &  & 5    & 30.85  & 14.82 \\
        \hline
        21 & \multirow{5}{*}{10} & -5   & 32.43  & 13.53 \\
        22 &  & -2.5 & 32.65  & 14.47 \\
        23 &  & 0    & 32.13  & 14.02 \\
        24 &  & 2.5  & 32.73  & 14.12 \\
        25 &  & 5    & 32.05  & 14.60 \\
        \hline
        Additional case number & Torso Angle ($x_1$) & D-ring $Z$ Position ($x_2$) & HIC15 & $a_{\text{T1,max}}$ [$\mathrm{m/s^2}$]\\
        \hline
        26 & -2.5 &  -5 & 24.28  & 13.98 \\
        27 &  2.5 & 0   & 27.54  & 13.43 \\
        \hline
    \end{tabular}
\end{table}

\end{document}